\documentclass{appolb}
\usepackage{cite,epsfig,psfig}

\def\beq{\begin{equation}}
\def\eeq{\end{equation}}
\def\beqar{\begin{eqnarray}}
\def\eeqar{\end{eqnarray}}
\def\barr#1{\begin{array}{#1}}
\def\earr{\end{array}}
\def\bfi{\begin{figure}}
\def\efi{\end{figure}}
\def\btab{\begin{table}}
\def\etab{\end{table}}
\def\bce{\begin{center}}
\def\ece{\end{center}}

\def\text{\textstyle}

\def\al{\alpha}

\def\De{\Delta}

\def\refeq#1{\mbox{eq.~(\ref{#1})}}
\def\refeqs#1{\mbox{eqs.~(\ref{#1})}}
\def\reffi#1{\mbox{Fig.~\ref{#1}}}

\def\citere#1{\mbox{Ref.~\cite{#1}}}
\def\citeres#1{\mbox{Refs.~\cite{#1}}}

\def\mathswitchr#1{\relax\ifmmode{\mathrm{#1}}\else$\mathrm{#1}$\fi}

\newcommand{\PW}{\mathswitchr W}
\newcommand{\PZ}{\mathswitchr Z}
\newcommand{\PA}{\mathswitchr A}

\newcommand{\PH}{\mathswitchr H}

\newcommand{\Ph}{\mathswitchr h}

\newcommand{\Pt}{\mathswitchr t}

\def\mathswitch#1{\relax\ifmmode#1\else$#1$\fi}

\newcommand{\MW}{\mathswitch {M_\PW}}

\newcommand{\MZ}{\mathswitch {M_\PZ}}

\newcommand{\Mt}{\mathswitch {m_\Pt}}
\newcommand{\MA}{\mathswitch {M_\PA}}

\newcommand{\scrs}{\scriptscriptstyle}
\newcommand{\sw}{\mathswitch {s_{\scrs\PW}}}

\newcommand{\cw}{\mathswitch {c_{\scrs\PW}}}

\newcommand{\GF}{\mathswitch {G_\mu}}
\newcommand{\gf}{\GF}

\def\tb{\tan\beta}

\newcommand{\CTb}{\cot \beta\hspace{1mm}}

\newcommand{\Sb}{\sin \beta\hspace{1mm}}
\newcommand{\SQb}{\sin^2\beta\hspace{1mm}}

\newcommand{\CQb}{\cos^2\beta\hspace{1mm}}

\newcommand{\CZb}{\cos 2\beta\hspace{1mm}}
\newcommand{\CQZb}{\cos^2 2\beta\hspace{1mm}}

\newcommand{\mh}{m_\Ph}
\newcommand{\mH}{m_\PH}
\newcommand{\mst}{m_{\tilde{t}}}
\newcommand{\delmst}{\Delta\mst}

\newcommand{\mste}{m_{\tilde{\Pt}_1}}
\newcommand{\mstz}{m_{\tilde{\Pt}_2}}

\newcommand{\MstL}{M_{\tilde{t}_L}}
\newcommand{\MstR}{M_{\tilde{t}_R}}

\newcommand{\Mtlr}{M_{t}^{LR}}

\newcommand{\Mtlrz}{\KL M_{t}^{LR}\KR^2}

\newcommand{\Mtlrv}{\KL M_{t}^{LR}\KR^4}

\newcommand{\Mtlrse}{\KL M_{t}^{LR}\KR^6}

\newcommand{\Mtlra}{\KL M_{t}^{LR}\KR^8}

\newcommand{\lmtmsms}{\KL\frac{\mtms^2}{\ms^2}\KR}

\newcommand{\mt}{\Mt}
\newcommand{\mtms}{\overline{m}_{\Pt}}

\newcommand{\mgl}{m_{\tilde{\mathrm{g}}}}

\newcommand{\Stop}{\tilde{t}}
\newcommand{\StopL}{\tilde{t}_L}
\newcommand{\StopR}{\tilde{t}_R}
\newcommand{\Stope}{\tilde{t}_1}
\newcommand{\Stopz}{\tilde{t}_2}

\newcommand{\Sbot}{\tilde{b}}

\newcommand{\tst}{\theta_{\tilde{\Pt}}}

\newcommand{\tsf}{\theta\kern-.20em_{\tilde{f}}}
\newcommand{\tsfp}{\theta\kern-.20em_{\tilde{f}\prime}}
\newcommand{\tsq}{\theta\kern-.15em_{\tilde{q}}}

\newcommand{\ms}{M_{\mathrm{S}}}
\newcommand{\msq}{m_{\tilde{q}}}

\newcommand{\Pe}{\phi_1}
\newcommand{\Pz}{\phi_2}

\newcommand{\lsim}
{\;\raisebox{-.3em}{$\stackrel{\displaystyle <}{\sim}$}\;}

\newcommand{\fea}{{\em FeynArts}}

\newcommand{\two}{{\em TwoCalc}}
\newcommand{\fh}{{\em FeynHiggs}}
\newcommand{\fhi}{\fh}
\newcommand{\fhf}{{\em Feyn\-Higgs\-Fast}}

\newcommand{\msbar}{$\overline{\rm{MS}}$}
\newcommand{\SU}{\mathrm {SUSY}}
\newcommand{\oa}{{\cal O}(\alpha)}

\newcommand{\oaas}{{\cal O}(\alpha\alpha_s)}
\newcommand{\cp}{{\cal CP}}
\newcommand{\wz}{\sqrt{2}}
\newcommand{\edz}{\frac{1}{2}}
\newcommand{\twol}{two-loop}
\newcommand{\onel}{one-loop}

\newcommand{\dst}{\Delta_{\tilde{t}}}

\newcommand{\dr}{\De\rho}

\newcommand{\KL}{\left(}
\newcommand{\KR}{\right)}
\newcommand{\KKL}{\left[}
\newcommand{\KKR}{\right]}
\newcommand{\KKKL}{\left\{}
\newcommand{\KKKR}{\right\}}

\newcommand{\VL}{\left( \begin{array}{c}}
\newcommand{\VR}{\end{array} \right)}
\newcommand{\ML}{\left( \begin{array}{cc}}
\newcommand{\MLd}{\left( \begin{array}{ccc}}
\newcommand{\MLv}{\left( \begin{array}{cccc}}
\newcommand{\MR}{\end{array} \right)}

\newcommand{\gev}{\,\, \mathrm{GeV}}

\newcommand{\BC}{\begin{center}}
\newcommand{\EC}{\end{center}}
\newcommand{\BE}{\begin{equation}}
\newcommand{\EE}{\end{equation}}
\newcommand{\BEA}{\begin{eqnarray}}
\newcommand{\BEAnn}{\begin{eqnarray*}}
\newcommand{\EEA}{\end{eqnarray}}
\newcommand{\EEAnn}{\end{eqnarray*}}
\newcommand{\non}{\nonumber}
\newcommand{\id}{{\rm 1\kern-.12em
\rule{0.3pt}{1.5ex}\raisebox{0.0ex}{\rule{0.1em}{0.3pt}}}}

\def\als{\alpha_s}

\def\hSi{\hat{\Sigma}}

\def\hSie{\hat{\Sigma}^{(1)}}
\def\hSiz{\hat{\Sigma}^{(2)}}

\def\draftdate{\relax}
\def\mda{\relax}
\def\mua{\relax}
\def\mla{\relax}
\def\draft{
\def\thtystars{******************************}
\def\sixtystars{\thtystars\thtystars}
\typeout{}
\typeout{\sixtystars**}
\typeout{* Draft mode!
         For final version remove \protect\draft\space in source file
*}
\typeout{\sixtystars**}
\typeout{}
\def\draftdate{\today}
\def\mua{\marginpar[\boldmath\hfil$\uparrow$]%
                   {\boldmath$\uparrow$\hfil}%
                    \typeout{marginpar: $\uparrow$}\ignorespaces}
\def\mda{\marginpar[\boldmath\hfil$\downarrow$]%
                   {\boldmath$\downarrow$\hfil}%
                    \typeout{marginpar: $\downarrow$}\ignorespaces}
\def\mla{\marginpar[\boldmath\hfil$\rightarrow$]%
                   {\boldmath$\leftarrow $\hfil}%
                    \typeout{marginpar:
$\leftrightarrow$}\ignorespaces}
\def\Mua{\marginpar[\boldmath\hfil$\Uparrow$]%
                   {\boldmath$\Uparrow$\hfil}%
                    \typeout{marginpar: $\Uparrow$}\ignorespaces}
\def\Mda{\marginpar[\boldmath\hfil$\Downarrow$]%
                   {\boldmath$\Downarrow$\hfil}%
                    \typeout{marginpar: $\Downarrow$}\ignorespaces}
\def\Mla{\marginpar[\boldmath\hfil$\Rightarrow$]%
                   {\boldmath$\Leftarrow $\hfil}%
                    \typeout{marginpar:
$\Leftrightarrow$}\ignorespaces}
\overfullrule 5pt
\oddsidemargin -15mm
\marginparwidth 29mm
}

\begin{document}

\thispagestyle{empty}

\null
\hfill KA-TP-4-1999\\
\null
\hfill DESY 99-043\\
\null
\hfill hep-ph/9903504\\
\vskip .8cm
\begin{center}
{\Large \bf Precision Analysis of the Masses\\[.5em]
of the Neutral Higgs Bosons in the MSSM
}
\vskip 2.5em
{\large
{\sc S.\ Heinemeyer$^a$, W.\ Hollik$^{b,c}$ and G.\ Weiglein$^c$}\\[1em]
{\normalsize \it $^a$ DESY Theorie, Notkestr. 85, 22603 Hamburg,
Germany}\\[.3em]
{\normalsize \it $^b$ Theoretical Physics Division, CERN, CH-1211
Geneva 23, Switzerland}\\[.3em]
{\normalsize \it $^c$ Institut f\"ur Theoretische Physik, Universit\"at
Karlsruhe,\\
D-76128 Karlsruhe, Germany}
}
\vskip 2em
\end{center} \par
\vskip 1.2cm
\vfil
\bce
{\bf Abstract} \par
\ece
The masses of the neutral $\cp$-even Higgs bosons in the Minimal
Supersymmetric Standard Model (MSSM) are predicted on the basis of 
explicit Feynman-diagrammatic calculations. The results, 
containing the complete diagrammatic \onel\ corrections, the leading
\twol\ corrections of $\oaas$ and further improvements taking
into account leading electroweak \twol\ and higher-order QCD
contributions, are discussed 
and compared with results obtained by renormalization
group calculations. Good agreement is found in the case of vanishing
mixing in the scalar top sector, while sizable deviations occur if 
scalar top mixing is taken into account.
By means of a Taylor expansion a compact approximation formula for
the mass of the lightest Higgs boson, $\mh$, is derived. The quality 
of the approximation in comparison with the full result is analyzed.
\par
\vskip 1cm
\null
\setcounter{page}{0}
\clearpage

\pagestyle{plain}
\newcount\eLiNe\eLiNe=\inputlineno\advance\eLiNe by -1
\title{PRECISION ANALYSIS OF THE MASSES\\
OF THE NEUTRAL HIGGS BOSONS IN THE MSSM%
\thanks{Talk given by G.~Weiglein at the Cracow Epiphany Conference on
Electron-Positron Colliders, Cracow, January 5--10, 1999.}
}
\author{Sven Heinemeyer$^a$, Wolfgang Hollik$^{b,c}$ and
Georg Weiglein$^c$
\address{$^a$ DESY Theorie, Notkestr. 85, 22603 Hamburg, Germany\\[.3em]
  $^b$ Theoretical Physics Division, CERN, CH-1211 Geneva 23, 
       Switzerland\\[.3em]
  $^c$ Institut f\"ur Theoretische Physik, Universit\"at Karlsruhe,\\
       D--76128 Karlsruhe, Germany
}}
\maketitle

\begin{abstract}
The masses of the neutral $\cp$-even Higgs bosons in the Minimal
Supersymmetric Standard Model (MSSM) are predicted on the basis of 
explicit Feynman-diagrammatic calculations. The results, 
containing the complete diagrammatic \onel\ corrections, the leading
\twol\ corrections of $\oaas$ and further improvements taking
into account leading electroweak \twol\ and higher-order QCD
contributions, are discussed 
and compared with results obtained by renormalization
group calculations. Good agreement is found in the case of vanishing
mixing in the scalar top sector, while sizable deviations occur if 
scalar top mixing is taken into account.
By means of a Taylor expansion a compact approximation formula for
the mass of the lightest Higgs boson, $\mh$, is derived. The quality 
of the approximation in comparison with the full result is analyzed.
\end{abstract}

\def\thefootnote{\arabic{footnote}}
\setcounter{footnote}{0}


\section{Introduction}
The search for the lightest Higgs boson provides a direct and very 
stringent test of Supersymmetry (SUSY) and is one of the main goals
at the present and the next generation of colliders. A precise
prediction of its mass, $\mh$, is inevitable for determining the
discovery and exclusion potential of LEP2 and the upgraded Tevatron in
this search and for analyzing the accessible MSSM parameter space.
If the MSSM Higgs boson exists, it will be detectable at the LHC and a
future linear collider (LC), and its mass will be measured at these machines
with high precision. The comparison of the MSSM prediction with 
the experimental value of $\mh$ will then allow a very sensitive 
test of the model. 
A precise knowledge of the mass of the heavier $\cp$-even Higgs boson,
$\mH$, will be important for
resolving the mass splitting between the $\cp$-even and -odd
Higgs-boson masses.

The mass of the lightest Higgs boson in the MSSM is restricted at the 
tree level to be smaller than the $\PZ$-boson mass, $\MZ$.
The dominant \onel\ corrections arise from the top and scalar-top sector 
via terms of the form $\GF \mt^4 \ln (\mste \mstz/\mt^2)$~\cite{mhiggs1l}. 
These results have been improved by performing a complete \onel\ calculation 
in the on-shell scheme~\cite{mhoneloop,mhiggsf1l,mhiggsf1lc}. 
Beyond \onel\ order renormalization group (RG)
methods have been applied in order to obtain leading logarithmic
higher-order
contributions~\cite{mhiggs1lrest,mhiggsRG1,mhiggsRG1a,mhiggsRG1b,mhiggsRG2}.
Furthermore the leading \twol\ QCD corrections have been calculated in the
effective potential method~\cite{mhiggsEffPotHH,mhiggsEffPotZ}.
Phenomenological analyses for the 
neutral $\cp$-even Higgs-boson masses have until recently been based
either on RG improved \onel\
calculations~\cite{mhiggsRG1,mhiggsRG1a,mhiggsRG2} or on the complete
Feynman-diagrammatic \onel\ on-shell
result~\cite{mhoneloop,mhiggsf1l,mhiggsf1lc}. The numerical results of
these approaches however differ by up to $20 \gev$ in $\mh$.

Recently the Feynman-diagrammatic result for the dominant \twol\
contributions of $\oaas$ to the masses of the neutral $\cp$-even 
Higgs bosons has become available~\cite{mhiggsletter}. By combining 
these contributions
with the complete \onel\ on-shell result~\cite{mhiggsf1l}, the
currently most precise result for $\mh$ based on diagrammatic
calculations is obtained~\cite{mhiggsletter2,mhiggslong}. It has 
been implemented into a Fortran program called \fh~\cite{feynhiggs}.
In the present paper the new Feynman-diagrammatic results are briefly 
summarized and compared with the results obtained by RG methods. 
Furthermore a compact analytical approximation formula~\cite{mhiggslle}
is discussed, which 
is derived from the full diagrammatic result by means of a Taylor expansion.


\section{Diagrammatic two-loop calculation of the masses of the neutral
$\cp$-even Higgs bosons}

The MSSM Higgs sector can be described with the help of two pa\-ra\-meters: 
$\tb = v_2/v_1$, the ratio of the two vacuum expectation values, and
$\MA$, the mass of the $\cp$-odd Higgs boson.
The tree-level predictions for the masses $\mh$ and $\mH$ of the 
neutral $\cp$-even Higgs bosons $h$ and $H$
are determined by diagonalizing the
tree-level mass matrix given in terms of the current eigenstates $\Pe$
and $\Pz$.
In the Feynman-diagrammatic approach the higher-order corrected 
Higgs-boson
masses are derived by determining the poles of the $h,H$-propagator
matrix whose inverse 
is given by
\BE
\left(\Delta_{\rm Higgs}\right)^{-1}
= - i \ML q^2 -  m_{\PH,{\rm tree}}^2 + \hSi_{H}(q^2) &  \hSi_{hH}(q^2) \\
     \hSi_{hH}(q^2) & q^2 -  m_{\Ph,{\rm tree}}^2 + \hSi_{h}(q^2) \MR,
\label{higgsmassmatrixnondiag}
\EE
where the $\hSi$ denote 
the renormalized Higgs-boson self-energies,
which can be decomposed according to 
\BE
\hSi_s = \hSie_s + \hSiz_s + \ldots , \quad s = h, H, hH,
\EE
into the contributions at \onel\ order, \twol\ order etc.

For the
\onel\ contributions to these self-energies, $\hSie_s(q^2)$,
we take the result of the complete one-loop on-shell calculation of
\citere{mhiggsf1l}. The agreement with the result obtained in 
\citere{mhoneloop} is better than $1 \gev$ for almost the whole MSSM
parameter space. 

The leading \twol\ corrections, $\hSiz_s(0)$, have been obtained in
Refs.\ \cite{mhiggsletter,mhiggsletter2,mhiggslong} by calculating the
$\oaas$ contribution  
of the $t,\Stop$-sector to the renormalized Higgs-boson self-energies
at zero external momentum from the Yukawa part of the theory.
The calculation has been performed in the on-shell scheme. It involves
a two-loop renormalization in the Higgs sector and a one-loop
renormalization in the scalar top sector of the MSSM. The calculations
have been performed using Dimensional Reduction (DRED)~\cite{dred},
which is necessary in order to preserve the relevant SUSY relations.
In deriving these results, use has been made of the computer-algebra
programs \fea~\cite{fea} (in which the relevant part of the MSSM
has been implemented) for generating the Feynman amplitudes, and
\two~\cite{two} for evaluating the two-loop diagrams and counterterm
contributions.

The results for the corrections in $\oaas$ are given in terms of the
SUSY parameters $\tb$, $\MA$, $\mu$, $\mgl$, $\mste$, $\mstz$, and $\tst$,
where $\mu$ denotes the Higgs-mixing parameter and $\mgl$ the mass of
the gluino. The mass eigenstates $\Stope$, $\Stopz$ and the mixing angle
$\tst$ in the scalar top sector are derived 
by diagonalizing the mass matrix of the scalar top quarks given in the
basis of the current eigenstates $\StopL$, $\StopR$.
The non-diagonal entry in the scalar quark mass matrix is proportional
to the mass of the quark and reads for the $\Stop$-mass matrix
$\mt \Mtlr = \mt (A_{\Pt} - \mu \CTb)$, 
where we have adopted the conventions as 
in~\citere{mhiggslong}. 
Due to the large value of $\mt$ these mixing effects are in general
non-negligible.

Inserting the contributions in $\oa$ and $\oaas$ into 
\refeq{higgsmassmatrixnondiag} and determining the poles of the 
$h,H$-propagator matrix yields the prediction for the masses of the
neutral $\cp$-even Higgs bosons. 
We have implemented two further corrections beyond $\oaas$ into the
prediction for $\mh$:
The first correction concerns leading QCD corrections beyond
\twol\ order, taken into account by using the \msbar\
top-quark mass
\BE
\mtms = \mtms(\mt) \approx \mt /\left(1 + \frac{4}{3\,\pi} \als(\mt)
\right)
\label{mtrun}
\EE
for the \twol\ contributions instead of the pole mass $\mt$.
The second one is the
leading \twol\ Yukawa correction of ${\cal O}(\gf^2 \mt^6)$,
taken over from the result obtained by RG
methods~\cite{mhiggsRG1a,ccpw}.

The results described above have been implemented into the Fortran
program \fh~\cite{feynhiggs}, which needs about $0.5$~seconds for the 
evaluation of $\mh$, $\mH$ on a 
Sigma station (Alpha CPU, 600 MHz) for one set of pa\-ra\-meters. 
As an additional constraint (besides the experimental bounds) on the
squark masses, the program also evaluates the contribution to $\De\rho$
arising from $\Stop/\Sbot$-loops up to $\oaas$~\cite{drhosuqcd}.
A value of $\De\rho$ outside the 
preferred region of 
$\dr^{\SU} \lsim 1 \cdot 10^{-3}$~\cite{delrhoexp} indicates 
experimentally disfavored $\Stop$- and $\Sbot$-masses.
The program \fh\ is available via the WWW page \\
{\tt http://www-itp.physik.uni-karlsruhe.de/feynhiggs}~.

\section{Numerical Results}

For the numerical evaluation we have chosen two values for
$\tb$, which are favored by GUT scenarios~\cite{su5so10}: 
$\tb = 1.6$ for
the $\mathrm{SU}(5)$ scenario and $\tb = 40$ for the $\mathrm{SO}(10)$ 
scenario. Other parameters are $\MZ = 91.187 \gev, \MW = 80.39 \gev, 
\GF = 1.16639 \, 10^{-5} \gev^{-2}, \als(\mt) = 0.1095$, and $\mt = 175 \gev$,
if not otherwise indicated. 
Further parameters are $\MA$, $\mgl$, $\mu$, and the $\mathrm{SU}(2)$ soft 
SUSY-breaking parameter $M (\equiv M_2)$.
The other gaugino mass parameter, $M_1$, is fixed via the
GUT relation $M_1 = (5\, \sw^2)/(3\, \cw^2)\, M$.
In the figures below we have chosen $\msq \equiv \MstL = \MstR$ for the
diagonal entries in the scalar top mass matrix.

\begin{figure}[ht]
\begin{center}
\epsfig{figure=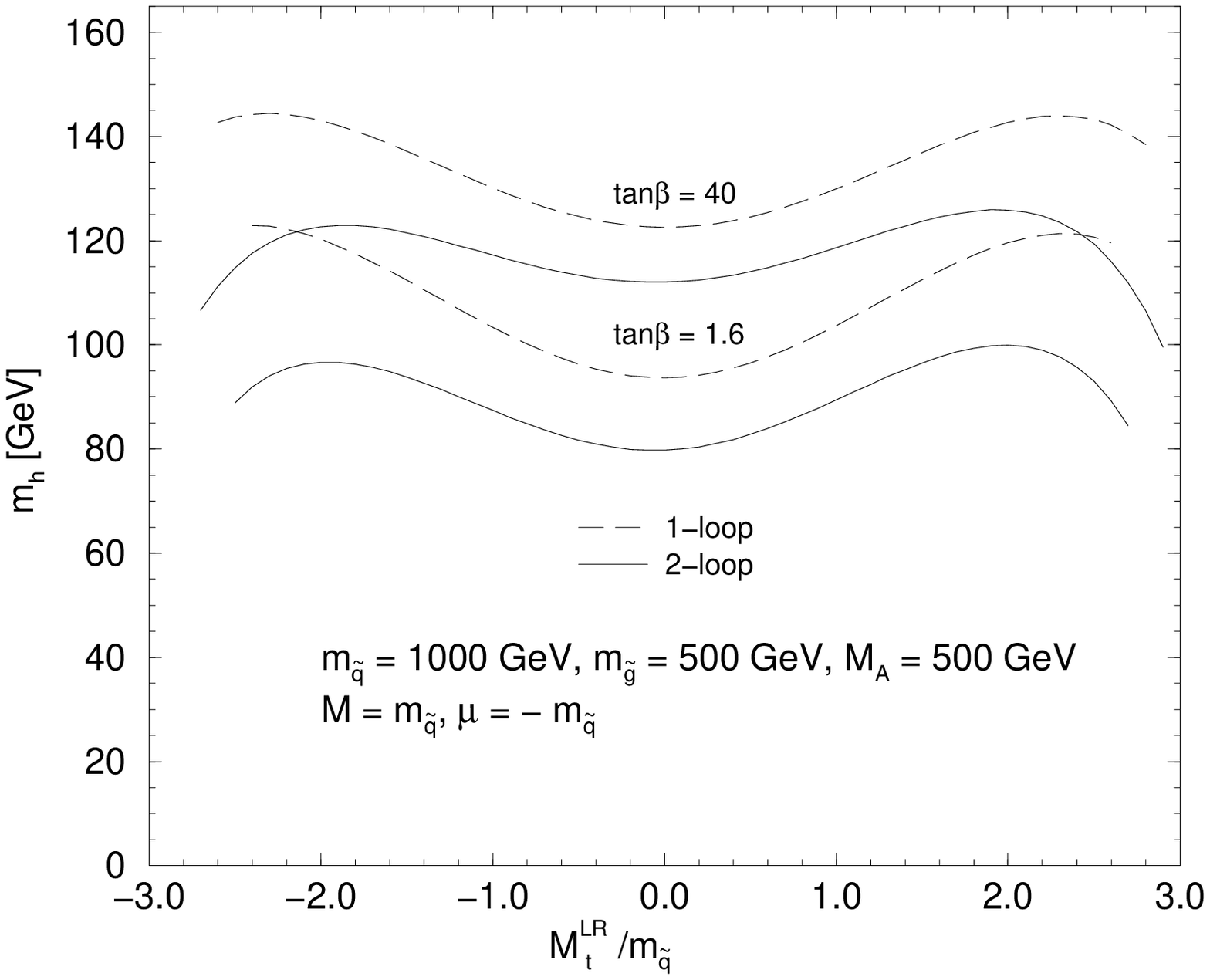,width=9.9cm,height=6.3cm}
\end{center}
\vspace{-0.5cm}
\caption[]{One- and \twol\ results for the mass of the lightest Higgs
boson $\mh$ as a function of $\Mtlr/\msq$ for two values of $\tb$.}
\label{fig:mh_MtLRdivmq}
\end{figure}

Fig.~\ref{fig:mh_MtLRdivmq} shows the result for $\mh$ obtained
from the diagrammatic \twol\ calculation 
as a function of $\Mtlr/\msq$, where $\msq$ is
fixed to $1000 \gev$. The \twol\ contributions give rise to a large
reduction of the \onel\ on-shell result by up to 20 GeV. A minimum in
the prediction for $\mh$ occurs around $\Mtlr/\msq = 0$, which
corresponds to the case of no mixing in the $\Stop$-sector. A maximum
in the \twol\ result for $\mh$ is reached for about $|\Mtlr/\msq|
\approx 2$, this case we refer to as `maximal mixing'. In the \twol\
result the maxima are shifted compared to their \onel\ values of about
$|\Mtlr/\msq| \approx 2.4$. Varying $\tb$ around the value $\tb = 1.6$
leads to a relatively large effect in $\mh$, while the effect of
varying $\tb$ around $\tb = 40$ is marginal. Different values of the 
gluino mass, $\mgl$, in the \twol\ contribution affect the prediction 
for $\mh$ by up to $\pm 2 \gev$ in the maximal-mixing scenario, while the 
effect is negligible in the no-mixing scenario.
Varying $M$, which enters via the non-leading \onel\ 
contributions, changes the value of $\mh$ by $\pm 1.5 \gev$. 
A more detailed analysis of the dependence of our results on the 
different SUSY parameters has been performed in \citere{mhiggslong}.

\begin{figure}[ht]
\begin{center}
\epsfig{figure=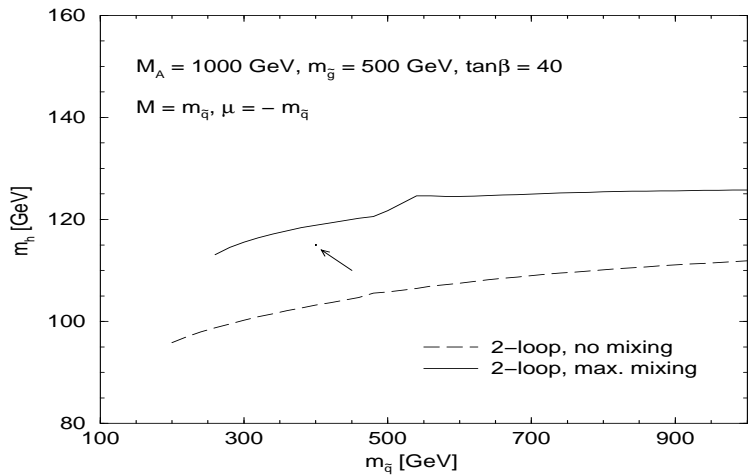,width=9.9cm,height=6.3cm}
\end{center}
\vspace{-0.5cm}
\caption[]{The \twol\ result for $\mh$ as a function of $\msq$
in the no-mixing and the maximal-mixing case.
The point marked by an arrow indicates the prospect for the experimental 
precision reached
by a future linear collider in the determination of $\mh$ and $\msq$
for the hypothetical values $\mh = 115$~GeV and $\msq = 400$~GeV.}
\label{fig:mh_mq}
\end{figure}

If the lightest Higgs boson and Supersymmetric particles will 
be found at the next generation of colliders, the experimental value
of $\mh$ will be measured with high accuracy and also the possible
range of the SUSY scale $\msq$ will in this case be constrained to a
small interval. At a high-luminosity
LC the prospect for the accuracy obtainable for these parameters
is $\De \mh = 0.05$~GeV and $\De \msq = 0.1 \%$. In \reffi{fig:mh_mq}
the \twol\ result for $\mh$ is shown as a function of $\msq$ in the
no-mixing and the maximal-mixing case. The parameter space in the
($\mh$, $\msq$) plane corresponding to the accuracy in $\mh$ and $\msq$
at the LC is indicated in the plot for the hypothetical central values
$\mh = 115$~GeV and $\msq = 400$~GeV. As can be seen from the plot, 
a precision determination of $\mh$ and $\msq$ will provide a very
sensitive consistency test of the model.

\begin{figure}[ht]
\begin{center}
\mbox{
\psfig{figure=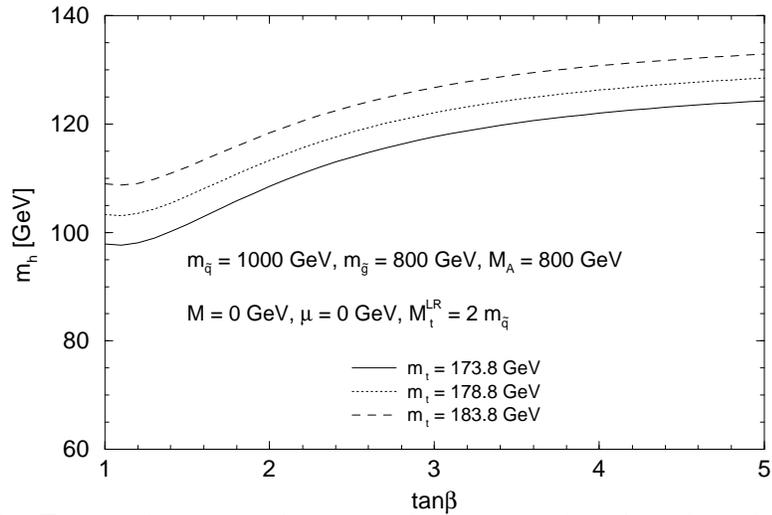,width=6.5cm,height=6.3cm,
                      bbllx=150pt,bblly=130pt,bburx=450pt,bbury=420pt}}
\end{center}
\vspace{0.2em}
\caption[]{
The maximally possible value for $\mh$ as a function of $\tb$ for 
$\msq = 1000$~GeV and three different values of $\mt$.
} 
\label{fig:mhmax}
\end{figure}

In order to determine the maximally possible value for $\mh$ within the
MSSM as a function of $\tb$, we have performed a parameter scan in which 
$\mgl, M, \mu, \MA$ and $\Mtlr$ have been varied for three values of
$\mt$ and fixed values of $\msq$ and $\tb$. Fig.~\ref{fig:mhmax} shows
the maximal Higgs-boson mass value in the range $\tb \leq 5$ for 
$\msq = 1000$~GeV (in Fig.~\ref{fig:mhmax} the
choice $M = \mu = 0$ has been made for simplicity; the change in $\mh$
when these parameters are chosen at their experimental lower bounds is
negligible). The upper bound is shown for the current experimental
value of the top-quark mass, $\mt = 173.8$~GeV, and for values which
are higher by one and two standard deviations, respectively. 
Our results confirm that
for the scenario with $\tb = 1.6$ practically the whole parameter space
of the MSSM can be covered at LEP2.
For slightly larger $\tb$ and maximal mixing, however, some parameter
space remains in which the Higgs boson could escape the detection at
LEP2.
For $\tb = 40$, on the other hand, the prediction for $\mh$ is at the
edge of the LEP2 range even in the no-mixing case. The full exploration 
of the MSSM parameter space for the scenario with large $\tb$ will be a
challenge for the upgraded Tevatron, the LHC, and the LC.

\section{Numerical comparison with the RG approach}

\begin{figure}[ht]
\vspace{-0.5cm}
\begin{center}
\mbox{
\psfig{figure=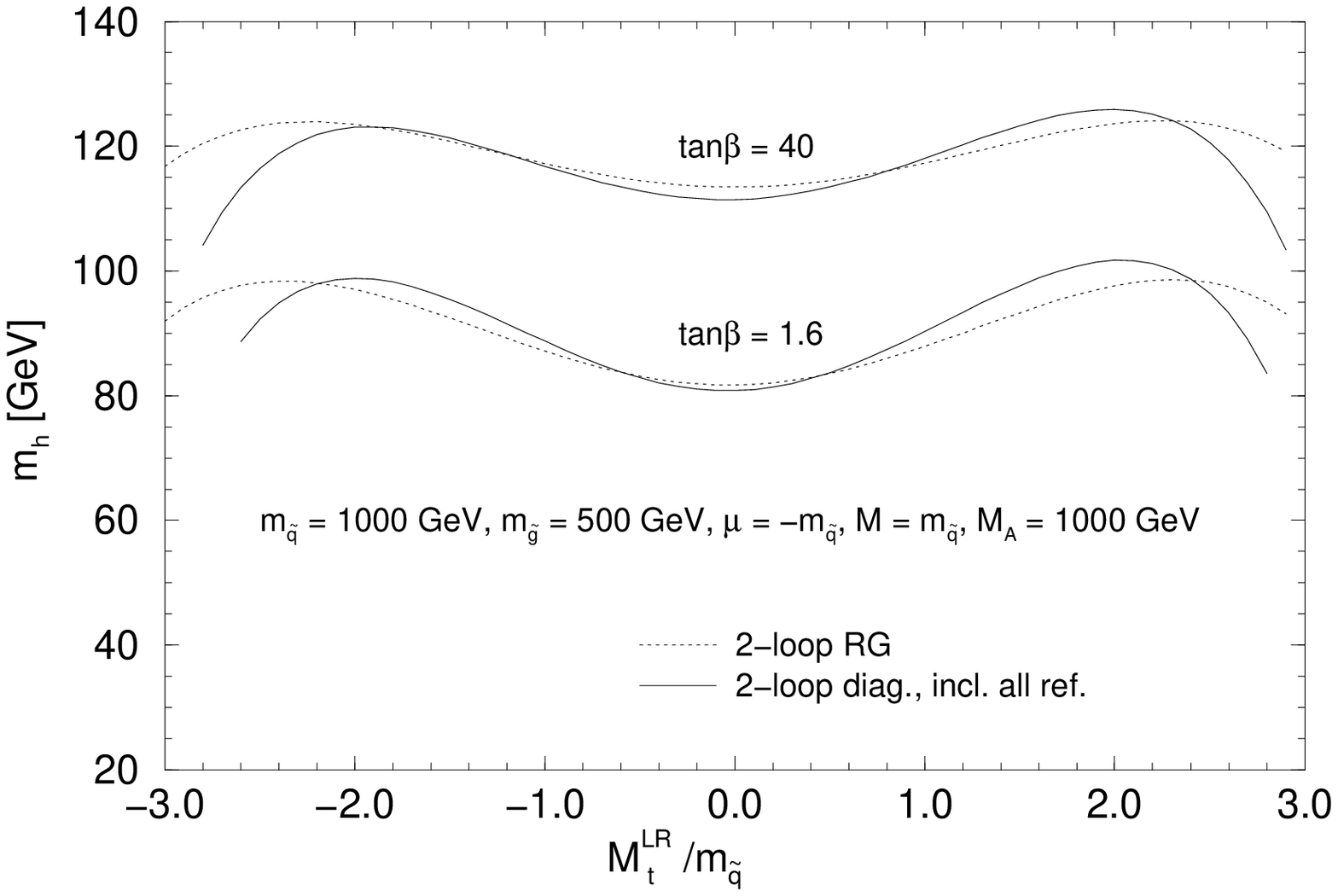,
       width=6.3cm,height=6.0cm,
       bbllx=150pt,bblly=130pt,bburx=450pt,bbury=420pt}}\\[1.5em]
\mbox{
\psfig{figure=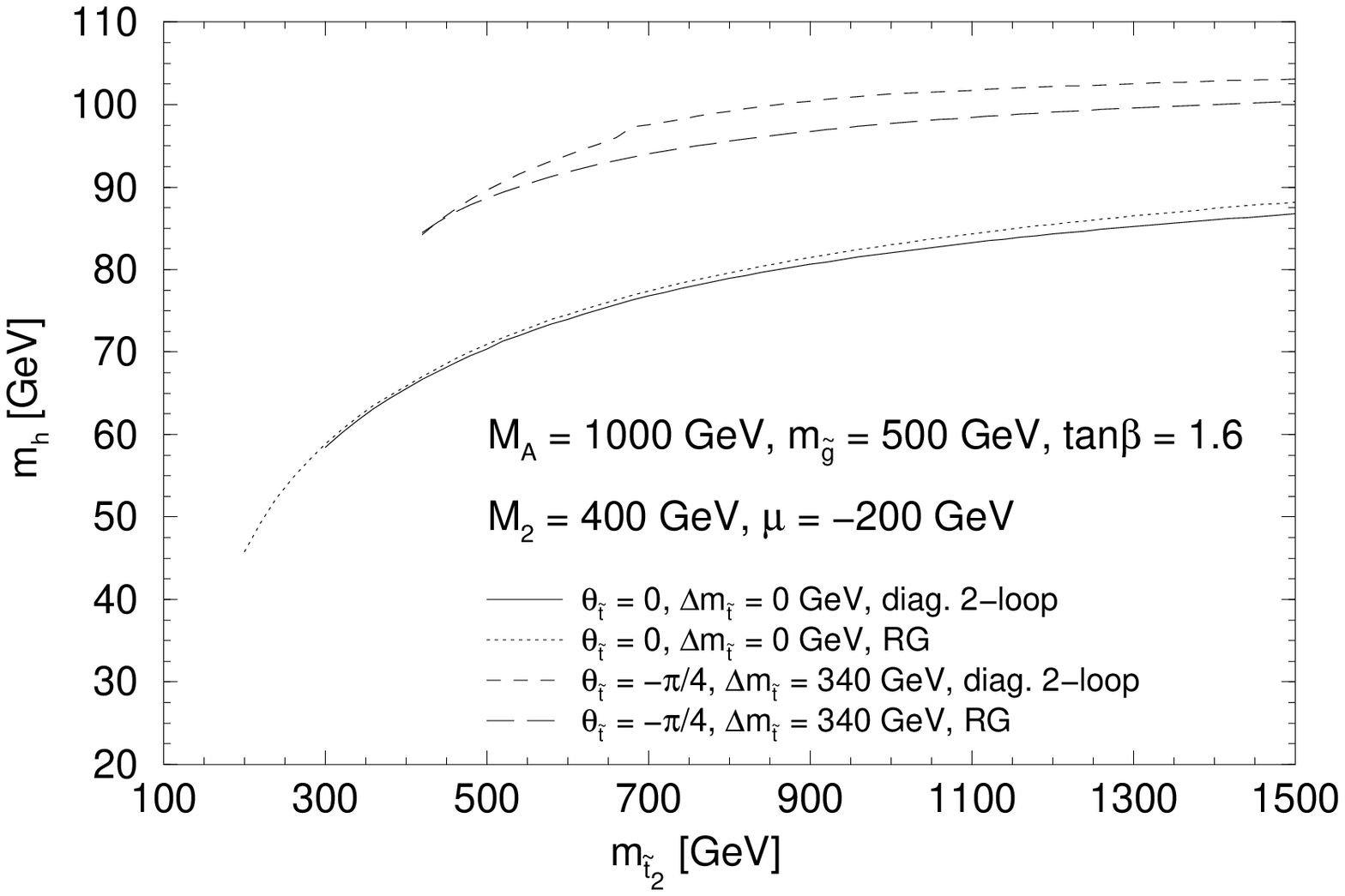,width=6.3cm,height=6.0cm,
                      bbllx=150pt,bblly=130pt,bburx=450pt,bbury=420pt}}
\end{center}
\vspace{0.5cm}
\caption[]{
Comparison between the Feynman-diagrammatic calculations and the
results obtained by renormalization group methods~\cite{mhiggsRG1b}.
In the upper plot the (unphysical) soft SUSY-breaking parameters of the
$\Stop$-mixing matrix are chosen as input, while
in the lower plot the physical $\Stop$-masses and the mixing angle
$\tst$ are the input parameters. For the curves with $\tst = 0$ in the
lower plot a mass difference $\delmst = 0 \gev$ is taken, whereas for
$\tst = -\pi/4$ we choose $\delmst = 340 \gev$, for
which the maximal Higgs-boson masses are achieved.
}
\label{fig:mh_RGVergleich}
\end{figure}

We now turn to the comparison of our diagrammatic results with the
predictions obtained via RG methods.
The upper plot of \reffi{fig:mh_RGVergleich} shows the prediction for
$\mh$ as a function of $\Mtlr/\msq$, corresponding to our diagrammatic
result and to the result obtained by RG methods~\cite{mhiggsRG1b}. In
the no-mixing case 
the diagrammatic result agrees well with the RG result. 
For non-vanishing
$\Stop$-mixing sizable deviations between the diagrammatic and the RG
results occur, which can reach $5 \gev$ for moderate mixing and become
very large for large values of $|\Mtlr/\msq|$. 
As already stressed above, the maximal
value for $\mh$ in the diagrammatic approach is reached for
$|\Mtlr/\msq| \approx 2$, whereas the RG results have a maximum at 
$|\Mtlr/\msq| \approx 2.4$, i.e. at the \onel\ value. Varying the value
of $\mgl$ in our result shifts the diagrammatic result relative to the RG
result (which does not contain the gluino mass as a parameter)
within $\pm 2 \gev$ in the region of large mixing.

In the upper plot of \reffi{fig:mh_RGVergleich} the results of our
diagrammatic on-shell 
calculation and the RG methods have been compared in terms of the
parameters $\MstL$, $\MstR$ and $\Mtlr$ of the $\Stop$-mixing matrix.
However, since the two approaches rely on different
renormalization schemes, the meaning of these (non-observable)
parameters is not precisely the same in the two approaches
starting from \twol\ order. 
In order to compare results obtained by different approaches making use
of different renormalization schemes, we find it preferable to compare
predictions for physical observables in terms of other observables
(instead of unphysical parameters). As a step into this direction we
compare in the lower plot of \reffi{fig:mh_RGVergleich} the diagrammatic
results and the RG 
results as a function of the physical mass $\mstz$ and with the mass
difference $\delmst = \mstz - \mste$ and the mixing angle $\tst$ as
parameters. 
In the context of the RG approach the running $\Stop$-masses, derived
from the $\Stop$-mass matrix, are considered as an approximation for
the physical masses.
As in the comparison performed above in terms of unphysical
parameters, in the lower plot of
\reffi{fig:mh_RGVergleich} very good agreement is found between 
the results of the two approaches in the case of vanishing
$\Stop$-mixing. For the maximal mixing angle $\tst = -\pi/4$ 
(and $\delmst = 340 \gev$, for which the maximal Higgs-boson masses are
achieved), however,
the diagrammatic result yields values for $\mh$ which are higher by
about $5 \gev$.

The upper bound on $\mh$ for a certain value of $\tb$ derived from our
diagrammatic results is thus higher in the low $\tb$ region by about
5~GeV than the upper bound derived previously from the RG results. As a
result, we find that the $\tb$ region which can fully be covered at
LEP2 and the upgraded Tevatron is significantly reduced compared to
previous studies.

\section{Compact approximation formula for $\mh$}

In order to extract the dominant contributions to $\mh$ from the rather 
complicated full result, we have derived by means of a 
Taylor expansion a short analytical approximation formula from the
diagrammatic \twol\ result~\cite{mhiggslle}. It can easily be
implemented into existing programs and allows a very fast numerical
evaluation. Since the most important contributions have been isolated
in this analytical formula, it is also helpful for a better qualitative
understanding of the source of the dominant corrections.

In deriving the formula the following approximations have been made:
\begin{itemize}
\item
The momentum dependence of the \onel\ and \twol\ self-energies
$\hSi_s$, $s = h, H, hH$ has been neglected in
\refeq{higgsmassmatrixnondiag}. 

\item 
The parameters $M$, $\mgl$ have been chosen according to 
$M = \mgl = \sqrt{\ms^2 - \mtms^2}$, where $\ms$ is given by
$$
\ms = \KKKL \begin{array}{l@{:\quad}l}
            \sqrt{\msq^2 + \mtms^2} & \MstL = \MstR = \msq \\
            \KKL \MstL^2 \MstR^2 + \mtms^2 (\MstL^2 + \MstR^2) +
                     \mtms^4 \KKR^\frac{1}{4} &
                                    \MstL \neq \MstR
            \end{array} \right. 
$$
\\[-2em]
\item
Contributions from the $t,\Stop$-sector up to the \twol\ level:\\
The main step of our approximations consists of a Taylor expansion of
the \onel\ and \twol\ contributions from the $t,\Stop$-sector in the 
parameter 
\BE
\dst = \frac{|\mt \Mtlr|}{\ms^2} =
       \frac{\mstz^2 - \mste^2}{\mstz^2 + \mste^2},
\EE
where terms proportional to $\MZ^2$ have been neglected in the
$\Stop$~mass matrix. For the \onel\ correction we have expanded up to
${\cal O}(\dst^8)$. We have
kept terms up to ${\cal O}(\MZ^4/\mt^4)$, while terms of 
${\cal O}(\MZ^2/\ms^2)$ have been neglected. For the \twol\ self-energies 
the expansion has been carried out up to 
${\cal O}(\dst^4)$.
We have furthermore used the approximation $\mu = 0$ in the
$\hSi_s(0)$. After extracting a common prefactor $(1/\sin^2\beta)$ we
have set otherwise $\Sb = 1$ in the non-logarithmic \onel\
contributions, while the full dependence on $\Sb$ is kept in the
logarithmic \onel\ and the \twol\ contributions. For a discussion of
these approximations see \citere{mhiggslle}.\\[-1em]
\item
For the \onel\ contributions from the other sectors of the MSSM the
leading logarithmic approximation has been used~\cite{mhiggs1lrest}. 
\\[-1em]
\item
Corrections beyond $\oaas$:\\
Leading contributions beyond $\oaas$ have been taken into account by
incorporating the leading \twol\ Yukawa
correction of ${\cal O}(\gf^2\mt^6)$ \cite{mhiggsRG1a,ccpw} and by
expressing the $t,\Stop$-contributions
through the \msbar\ top-quark mass $\mtms$ instead of the pole mass
$\mt$ according to \refeq{mtrun}.
\\[-1em]
\end{itemize}

The approximation formula for $\mh^2$ is obtained by inserting the
described approximations for the 
\onel\ and \twol\ self-energies $\hSi_s$ into the mass matrix 
\refeq{higgsmassmatrixnondiag}. The diagonalization of the mass matrix
incorporates contributions 
to $\mh^2$ that are formally of higher order but are
non-negligible in general. For large $\MA$ these higher-order
contributions are suppressed by inverse powers of $\MA$. Therefore it
is possible for $\MA \gg \MZ$ to perform an expansion in the loop
order, leading to a very compact formula for $\mh^2$ of the form
\BE
\label{mh2lle}
\mh^2 = \mh^{2,{\rm tree}} + \De\mh^{2,\al,t/\Stop}
         + \De\mh^{2,\al,{\rm rest}}
         + \De\mh^{2,\al\als} + \Delta\mh^{2,\al^2} .
\EE
The tree-level prediction and the one-loop contribution from the 
$t,\Stop$-sector are given by
\BEA
\label{mh2tree}
\lefteqn{
\mh^{2,{\rm tree}} = \edz \KKL \MA^2 + \MZ^2 
         - \sqrt{(\MA^2 + \MZ^2)^2 - 4 \MZ^2 \MA^2 \CQZb} \KKR , } \\
\lefteqn{ 
\De\mh^{2,\al,t/\Stop} = \frac{\gf\wz}{\pi^2}\; \mtms^4
      \Biggl[ \log\lmtmsms
             \Biggl\{ - \frac{3}{2}
                     - \frac{3}{4} \frac{\MZ^2}{\mtms^2} \CZb
                     - \frac{\MZ^4}{\mtms^4} \Lambda \CQZb \non }\\
 && {}               - \frac{\MZ^2}{\MA^2} \CQb \CZb
                       \left( 6 + \frac{3}{2} \frac{\MZ^2}{\mtms^2}
                           (1 - 4 \SQb) 
                      - \frac{\MZ^4}{\mtms^4} 8 \Lambda \CZb \SQb
                       \right) \Biggr\} \non \\
 && {}      +\Bigg\{  \frac{1}{4} \frac{\MZ^2}{\mtms^2}
                     -\frac{11}{80} \frac{\MZ^4}{\mtms^4}
           + \frac{\Mtlrz}{\ms^2}
                    \KL \frac{3}{2} - \edz \frac{\MZ^2}{\mtms^2}
                       - \frac{3}{4} \frac{\mtms^2}{\ms^2} \KR \non\\
 && {}            +\frac{\Mtlrv}{\ms^4}
                    \KL -\frac{1}{8} + \edz \frac{\mtms^2}{\ms^2}
                        -\frac{3}{8} \frac{\mtms^4}{\ms^4}
                    \KR \non\\
 && {}     +\frac{\Mtlrse}{\ms^6}
                    \KL -\frac{3}{40} \frac{\mtms^2}{\ms^2}
                        +\frac{3}{10} \frac{\mtms^4}{\ms^4}
                        -\frac{1}{4} \frac{\mtms^6}{\ms^6} \KR \non\\
 && {}            +\frac{\Mtlra}{\ms^8}
                    \KL -\frac{3}{56}\frac{\mtms^4}{\ms^4}
                        +\frac{3}{14}\frac{\mtms^6}{\ms^6}
                        -\frac{3}{16}\frac{\mtms^8}{\ms^8} \KR 
                        \Bigg\} \times \non \\
 && {}        \KL 1 + 4 \frac{\MZ^2}{\MA^2} \CQb \CZb \KR \Biggr] ,
\label{mh2onelooptop}
\EEA
where $\Lambda = \left(\frac{1}{8} - \frac{1}{3} \sw^2 + \frac{4}{9}
       \sw^4 \right)$, $\sw^2 = 1 - \frac{\MW^2}{\MZ^2}$.

The dominant \twol\ contribution of $\oaas$ to $\mh^2$ reads:
\BEA
\De\mh^{2,\al\als} &=&
    - \frac{\gf\wz}{\pi^2} \frac{\als}{\pi}\; \mtms^4
      \Biggl[ 4 + 3 \log^2\lmtmsms + 2 \log\lmtmsms
             -6 \frac{\Mtlr}{\ms} \non \\
 && {}  
     - \frac{\Mtlrz}{\ms^2} \KKKL 3 \log\lmtmsms +8 \KKKR \non \\
 && {} +\frac{17}{12} \frac{\Mtlrv}{\ms^4} \Biggr]
         \KL 1 + 4 \frac{\MZ^2}{\MA^2} \CQb \CZb \KR .
\label{mh2twolooptop}
\EEA
For the \onel\ contribution from the other sectors of the MSSM, 
$\De\mh^{2,\al,{\rm rest}}$, and the leading two-loop Yukawa correction, 
$\De\mh^{2,\al^2}$, which are numerically less important than the 
contributions given above, we refer to \citere{mhiggslle}. 
In the contributions from the $t,\Stop$-sector at \onel\ and \twol\ order,
\refeqs{mh2onelooptop} and (\ref{mh2twolooptop}), we have included
correction factors of ${\cal O}(\MZ^2/\MA^2)$. In this way the compact
formula~(\ref{mh2lle}) gives a reliable approximation for $\MA$ values
down to at least $\MA = 200 \gev$. The approximation formula for the 
general case of $\MA$ is given in \citere{mhiggslle}. 

The contribution of $\oaas$ given in \refeq{mh2twolooptop} can
be compared with analytical formulas derived via the \twol\ effective
potential approach for the case of no mixing in the
$\Stop$~sector~\cite{mhiggsEffPotHH}
and via RG methods~\cite{mhiggsRG1a,mhiggsRG2}. The leading
term $\sim\log^2(\mtms^2/\ms^2)$ agrees with the results in
\citeres{mhiggsRG1a,mhiggsRG2,mhiggsEffPotHH}. The subleading term for
vanishing $\Stop$-mixing $\sim\log(\mtms^2/\ms^2)$ agrees with the
result of the \twol\ effective potential approach~\cite{mhiggsEffPotHH}
and the result of the \twol\ RG
calculation~\cite{mhiggsEffPotHH,mhiggsRG2},
but differs from the RG improved \onel\
result~\cite{mhiggsRG1a,mhiggsRG2}. The term
$\sim\log(\mtms^2/\ms^2)(\Mtlr/\ms)^2$ for non-vanishing $\Stop$-mixing
differs from the result given in~\citere{mhiggsRG1a,mhiggsRG2}.
All other terms of $\oaas$ are new.
The term $\sim\Mtlr/\ms$ shows that the result for $\mh$ is not
symmetric in $\pm\Mtlr$.
The good numerical agreement with the RG results in the case of no
mixing in the $\Stop$~sector can qualitatively be understood by noting
that in the no-mixing case the leading term in both approaches agrees,
while for the corrections proportional to powers of $\Mtlr/\ms$
deviations occur already in the leading contribution.

\begin{figure}[ht]
\begin{center}
\epsfig{figure=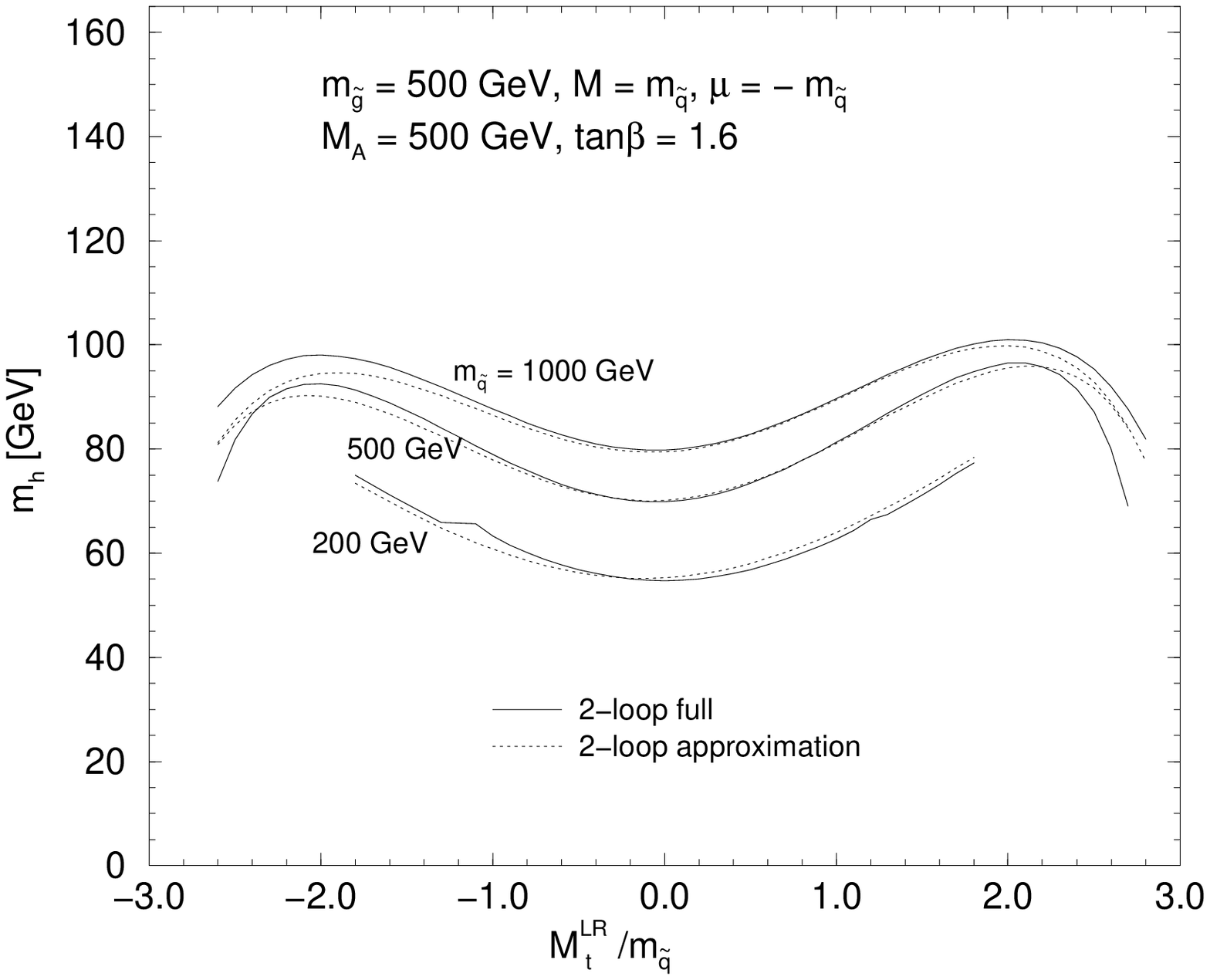,
           width=9.9cm,height=6.3cm}\\[1em]
\epsfig{figure=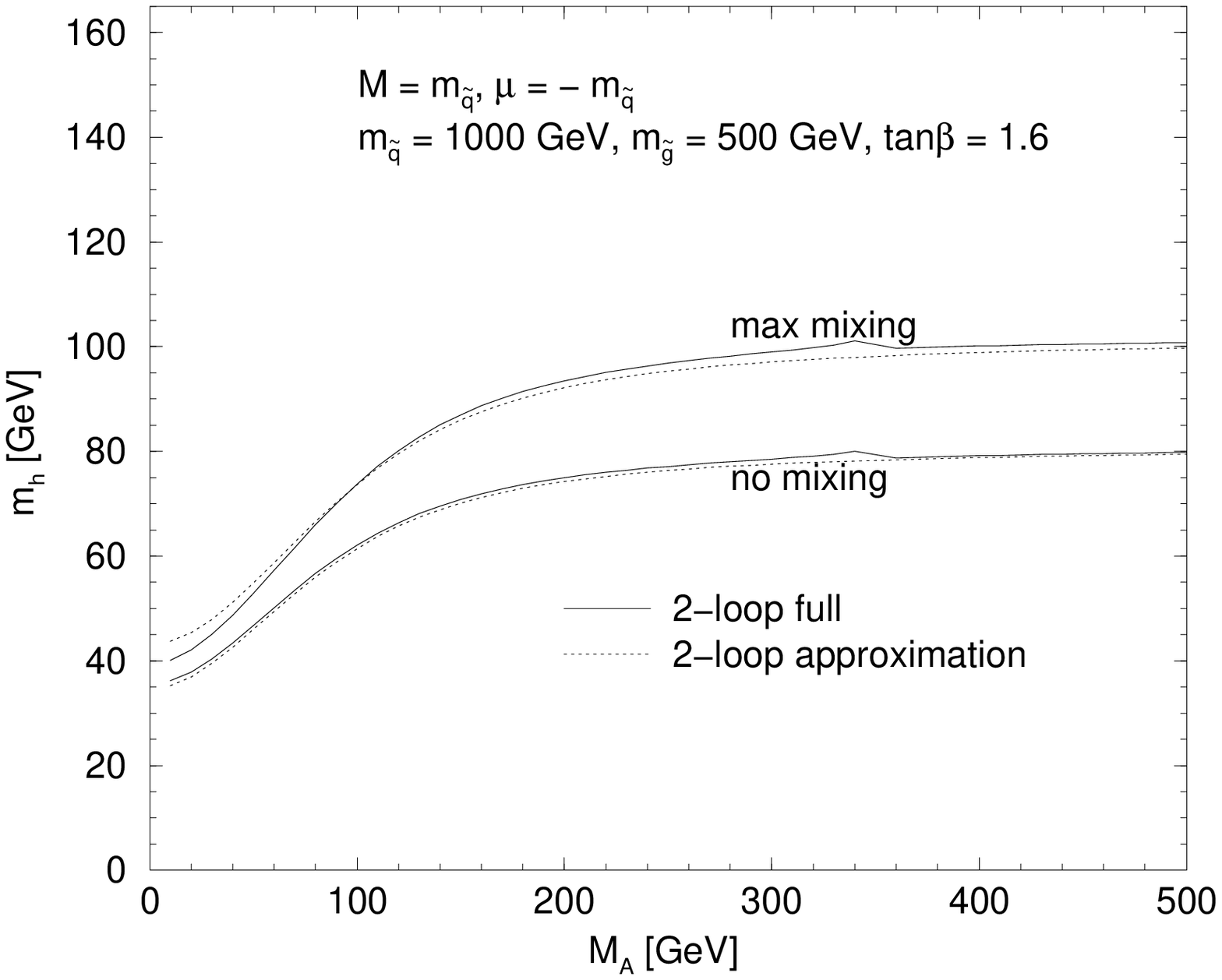,width=9.9cm,height=6.3cm}
\end{center}
\vspace{-0.5cm}
\caption[]{
Comparison between the 
approximation formula and the full diagrammatic result for $\mh$.
}
\label{fig:mhlle}
\end{figure}

The compact approximation formula has been implemented into \fh, in
order to allow a direct comparison between the full result and the 
approximation, and is also provided as a separate program called 
\fhf. The improvement in the speed of the evaluation with \fhf\
compared to \fhi\ is about a factor of $3 \times 10^{4}$.
In \reffi{fig:mhlle} the approximation formula is compared with
the full result (in the lower plot the formula for general $\MA$ is
applied). The compact formula approximates the full result better than about 
$2 \gev$ for most parts of the MSSM parameter space. Larger deviations 
can occur for $|\Mtlr/\msq| > 2$.

\bigskip

G.W.\ thanks M.~Je\.{z}abek and the other organizers of the
Epiphany Conference for the invitation, the excellent organization and
the pleasant atmos\-phere during the Conference.


\begin{thebibliography}{99}

\newcommand{\anp}[3]{{\sl Ann.~Phys.} {\bf #1} (19#2) #3}
\newcommand{\app}[3]{{\sl Acta~Phys.~Pol.} {\bf #1} (19#2) #3}
\newcommand{\cmp}[3]{{\sl Commun. Math. Phys.} {\bf #1} (19#2) #3}
\newcommand{\cpc}[3]{{\sl Comp. Phys. Commun.} {\bf #1} (19#2) #3}
\newcommand{\fp}[3]{{\sl Fortschr. Phys.} {\bf #1} (19#2) #3}
\newcommand{\ijmp}[3]{{\sl Int. J. Mod. Phys.} {\bf #1} (19#2) #3}
\newcommand{\jetp}[3]{{\sl JETP} {\bf #1} (19#2) #3}
\newcommand{\jetpl}[3]{{\sl JETP Lett.} {\bf #1} (19#2) #3}
\newcommand{\jmp}[3]{{\sl J. Math. Phys.} {\bf #1} (19#2) #3}
\newcommand{\mpl}[3]{{\sl Mod. Phys. Lett.} {\bf #1} (19#2) #3}
\newcommand{\nc}[3]{{\sl Nuovo Cimento} {\bf #1} (19#2) #3}
\newcommand{\ncl}[3]{{\sl Nuovo Cimento Lett.} {\bf #1} (19#2) #3}
\newcommand{\nim}[3]{{\sl Nucl. Instr. Meth.} {\bf #1} (19#2) #3}
\newcommand{\np}[3]{{\sl Nucl. Phys.} {\bf #1} (19#2)~#3}
\newcommand{\npB}[3]{{\sl Nucl. Phys.} {\bf B #1} (19#2)~#3}
\newcommand{\nphbps}[3]{{\sl Nucl. Phys.} {\bf B} {\it (Proc. Suppl.)}
{\bf #1B} (19#2) #3}
\newcommand{\plB}[3]{{\sl Phys. Lett.} {\bf B #1} (19#2) #3}
\newcommand{\prD}[3]{{\sl Phys. Rev.} {\bf D #1} (19#2) #3}
\newcommand{\prl}[3]{{\sl Phys. Rev. Lett.} {\bf #1} (19#2) #3}
\newcommand{\pl}[3]{{\sl Phys. Lett.} {\bf #1} (19#2) #3}
\newcommand{\ptp}[3]{{\sl Prog. Theo. Phys.} {\bf #1} (19#2) #3}
\newcommand{\sptp}[3]{{\sl Suppl. Prog. Theo. Phys.} {\bf #1} (19#2) #3}
\newcommand{\sjnp}[3]{{\sl Sov. J. Nucl. Phys.} {\bf #1} (19#2) #3}
\newcommand{\zp}[3]{{\sl Z. Phys.} {\bf #1} (19#2) #3}
\newcommand{\vj}[4]{{\sl #1~}{\bf #2} (19#3) #4}
\newcommand{\ej}[3]{{\bf #1} (19#2) #3}
\newcommand{\vjs}[2]{{\sl #1~}{\bf #2}}



\bibitem{mhiggs1l} H.~Haber and R.~Hempfling,
                   {\em Phys. Rev. Lett.} {\bf 66} (1991) 1815;\\
                   Y.~Okada, M.~Yamaguchi and T.~Yanagida,
                   {\em Prog. Theor. Phys.} {\bf 85} (1991) 1;\\
                   J.~Ellis, G.~Ridolfi and F.~Zwirner,
                   {\em Phys. Lett.} {\bf B 257} (1991) 83; 
                   {\em Phys. Lett.} {\bf B 262} (1991) 477;\\
                   R.~Barbieri and M.~Frigeni,
                   {\em Phys. Lett.} {\bf B 258} (1991) 395.

\bibitem{mhoneloop} P.~Chankowski, S.~Pokorski and J.~Rosiek,
                    {\em Nucl. Phys.} {\bf B 423} (1994) 437.

\bibitem{mhiggsf1l} A.~Dabelstein, 
                    {\em Nucl. Phys.} {\bf B 456} (1995) 25;
                    {\em Z. Phys.} {\bf C 67} (1995) 495.

\bibitem{mhiggsf1lc} J.~Bagger, K.~Matchev, D.~Pierce and R.~Zhang,
                 {\em Nucl. Phys.} {\bf B 491} (1997) 3.

\bibitem{mhiggs1lrest} H.~Haber and R.~Hempfling,
                       {\em Phys. Rev.} {\bf D 48} (1993) 4280.

\bibitem{mhiggsRG1}   J.~Casas, J.~Espinosa, M.~Quir\'os and A.~Riotto,
                      {\em Nucl. Phys.} {\bf B 436} (1995) 3,
                      E: {\em ibid.} {\bf B 439} (1995) 466.

\bibitem{mhiggsRG1a}  M.~Carena, J.~Espinosa, M.~Quir\'os and C.~Wagner,
                      {\em Phys. Lett.} {\bf B 355} (1995) 209.

\bibitem{mhiggsRG1b}  M.~Carena, M.~Quir\'os and C.~Wagner,
                      {\em Nucl. Phys.} {\bf B 461} (1996) 407.

\bibitem{mhiggsRG2} H.~Haber, R.~Hempfling and A.~Hoang,
                    {\em Z. Phys.} {\bf C 75} (1997) 539.

\bibitem{mhiggsEffPotHH} R.~Hempfling and A.~Hoang,
                       {\em Phys. Lett.} {\bf B 331} (1994) 99.

\bibitem{mhiggsEffPotZ} R.-J.~Zhang, {\em Phys. Lett.} {\bf B 447} (1999) 89.

\bibitem{mhiggsletter} S.~Heinemeyer, W.~Hollik and G.~Weiglein,
                       {\em Phys. Rev.} {\bf D 58} (1998) 091701.

\bibitem{mhiggsletter2} S.~Heinemeyer, W.~Hollik and G.~Weiglein,
                        {\em Phys. Lett.} {\bf B 440} (1998) 296.

\bibitem{mhiggslong} S.~Heinemeyer, W.~Hollik and G.~Weiglein,
                     CERN-TH/98-405, hep-ph/9812472,
                     to appear in {\em Eur. Phys. Jour.} {\bf C}.

\bibitem{feynhiggs} S.~Heinemeyer, W.~Hollik and G.~Weiglein,
                    CERN-TH/98-389, hep-ph/9812320.

\bibitem{mhiggslle} S.~Heinemeyer, W.~Hollik and G.~Weiglein,
                    CERN-TH/99-74, hep-ph/9903404.

\bibitem{dred}
W.~Siegel, \plB{84}{79}{193};\\
D.~Capper, D.~Jones and P.~van Nieuwenhuizen, \npB{167}{80}{479}.

\bibitem{fea}
J.~K\"ublbeck, M.~B\"ohm and A.~Denner, \cpc{60}{90}{165}.

\bibitem{two}
G.~Weiglein, R.~Scharf and M.~B\"ohm, \npB{416}{94}{606}.

\bibitem{ccpw} 
M.~Carena, P.~Chankowski, S.~Pokorski and C.~Wagner,
\plB{441}{98}{205}.

\bibitem{drhosuqcd} A.~Djouadi, P.~Gambino, S.~Heinemeyer, W.~Hollik,
                     C.~J\"unger and G.~Weiglein,
                     {\em Phys. Rev. Lett.} {\bf 78} (1997) 3626;
                     {\em Phys. Rev.} {\bf D 57} (1998) 4179.

\bibitem{delrhoexp} G.~Altarelli, hep-ph/9811456.

\bibitem{su5so10} M.~Carena, S.~Pokorski and C.~Wagner,
                  {\em Nucl. Phys.} {\bf B406} (1993) 59;\\
                  W.~de Boer et al.,
                  {\em Z. Phys.} {\bf C 71} (1996) 415.

\end{thebibliography}
\end{document}